\documentclass[conference,a4paper]{IEEEtran}%
\usepackage[utf8]{inputenc}
\usepackage{graphicx}
\usepackage{tabularx}
\usepackage{amsmath}
\usepackage{amssymb}
\usepackage{stackengine}
\usepackage{makecell}
\usepackage{multirow}
\usepackage{cite}
\usepackage{array}
\usepackage{bbding}
\usepackage{pifont}
\usepackage{wasysym}
\usepackage{comment}
\usepackage[font=footnotesize]{caption}

\usepackage{subcaption}
\usepackage[noend]{algpseudocode}
\usepackage{color}
\usepackage{booktabs}
\usepackage{mathtools}
\usepackage{dirtytalk}
\usepackage{etoolbox}
\usepackage{tikz}
\usepackage{algorithm,algpseudocode}
\usepackage{url}
\usepackage{float}
\usepackage{stfloats}

\definecolor{headerblue}{RGB}{30, 60, 114}
\definecolor{rowgray}{RGB}{240, 242, 245}

\renewcommand{\arraystretch}{1.5}
\title{Cognitive Energy Management: Concept, Framework, and Demonstration in Smart Port Energy Systems}

\makeatletter
\newcommand{\linebreakand}{%
  \end{@IEEEauthorhalign}
  \hfill\mbox{}\par
  \mbox{}\hfill\begin{@IEEEauthorhalign} \vspace{-10cm}
}
\makeatother
\author{
  \IEEEauthorblockN{Hafiz Majid Hussain}
  \IEEEauthorblockN{ \textit{School of Energy Systems   }}
  \IEEEauthorblockA{
       \textit{ Lappeenranta-Lahti } \\ \textit {University of Technology}\\
    Lappeenranta, Finland \\
    Majid.hussain@lut.fi
   }
  \and
  \IEEEauthorblockN{Wajiha Samar}
  \IEEEauthorblockA{ \textit {Department of Electrical}\\ \textit {and Energy Engineering}\\
   \textit{University of Vaasa,}\\
   Vaasa , Finland \\
   X9504085@student.uwasa.fi  }

 \and
 \IEEEauthorblockN{Juha Haakana}
  \IEEEauthorblockN{ \textit{School of Energy Systems   }}
  \IEEEauthorblockA{
      \textit{ Lappeenranta-Lahti } \\ \textit {University of Technology}\\
    Lappeenranta, Finland \\
    Juha.Haakana@lut.fi
   }
} 

\makeatletter
\patchcmd{\@maketitle}
  {\addvspace{0.5\baselineskip}\egroup}
  {\addvspace{-1\baselineskip}\egroup}
  {}
  {}
\makeatother

\begin{document}  
\maketitle
\begin{abstract}
Modern energy management systems, even within advanced energy internet (EI) infrastructures, remain fundamentally reactive, optimization-bound, and incapable of reasoning about context, intent, or uncertainty. While the EI paradigm has established a powerful cyber-physical architecture for interconnecting distributed energy resources via software-defined packetized networks, the question of how such systems should think, adapt, and govern energy decisions intelligently remains an open challenge. This paper introduces cognitive energy management (CEM); a new conceptual framework that addresses this gap by redefining how energy systems perceive, reason, learn, and act within complex operational environments. Grounded in the EI cyber-physical foundation, CEM extends beyond conventional optimization by embedding goal-directed reasoning and continuous adaptation into the energy management loop, positioning itself as the cognitive governance layer of EI-based infrastructures. We formally define CEM, distinguish it from rule-based and optimization-based paradigms through structured comparison, and articulate its core architectural layers. To demonstrate the framework's practical value, we develop a toy problem grounded in smart port energy management; one of the most operationally demanding EI node environments in modern infrastructure. Specifically, we model a predictive vessel turnaround scenario in which a CEM-enabled system plans energy procurement, storage pre-charging, and load scheduling across a six-hour operational horizon. The demonstration illustrates how CEM moves the EI beyond feasibility-seeking toward intelligent, anticipatory energy governance.
\end{abstract}
\begin{IEEEkeywords}
Energy internet, Cyber-physical, Cognitive energy management, Smart ports,  Software-defined packetized networks
\end{IEEEkeywords}
\section{Introduction}
The global energy landscape is undergoing a profound transformation. Driven by the rapid integration of renewable energy sources, the proliferation of distributed energy resources (DERs), and the urgent imperative of decarbonization, energy systems are growing not only in scale but in operational complexity \cite{onsomu2024comprehensive}. According to recent analyses, clean power generated more than 40\% of the world's electricity in 2024, with renewable capacity continuing to expand at an unprecedented pace \cite{weforum}. Yet, with greater renewable penetration comes greater volatility; energy production becomes intermittent, demand becomes harder to predict, and the decisions that govern energy flows become increasingly difficult for conventional systems to handle \cite{weforum1}.

In response to this complexity, the energy internet (EI) has emerged as a transformative paradigm. Defined as a cyber-physical system in which physical energy infrastructures and distributed renewable energy resources are interconnected and managed via a software-defined cyber energy network using packetized energy management techniques, the EI fundamentally reimagines how energy is generated, routed, and consumed \cite{hussain2020energy}. By integrating three tightly coupled layers: energy, communication, and information, the EI provides the architectural foundation for a new generation of intelligent, distributed, and responsive energy systems. Its core technologies, including energy routers, sub-energy routers, peer-to-peer trading platforms, and intelligent energy management software, represent a significant leap beyond the limitations of the conventional smart grid \cite{hussain2020energy}.
However, a critical gap remains. While the EI defines how energy systems are structured and interconnected, it does not fully resolve how they should think. The intelligent energy management (IEM) component envisioned within the EI framework still largely relies on optimization-based approaches; formulating energy management as a mathematical problem to be solved, given a fixed set of objectives and constraints \cite{hussain2020energy, graa2020review,hussain2024crossover, hussain2021packetized}. Such approaches are powerful within their assumptions, but they are fundamentally reactive: they respond to what is known, optimize for what is defined, and fail gracefully when faced with context they were not designed to anticipate. As explicitly noted in prior work, ``better and smarter energy management strategies must be employed for the optimal scheduling of energy resources" within EI-based systems \cite{hussain2020energy}, a challenge that remains open.

This is precisely the gap that motivates the introduction of cognitive energy management (CEM). Unlike conventional energy management systems: whether rule-based, model-driven, or optimization-bound, CEM approaches energy governance the way a skilled human operator would: by perceiving the operational context, reasoning about intent and uncertainty, learning from past experience, and acting with anticipation rather than mere reaction. CEM is not simply the application of machine learning to energy systems. Machine learning provides pattern recognition; cognition provides understanding. The distinction is fundamental; an ML model trained on historical data can predict tomorrow's demand, but a cognitive system can reason about why tomorrow's demand will differ, adapt its plan accordingly, and explain its decision in terms a human operator can validate. In this sense, CEM draws on cognitive computing principles \cite{safari2024energy}, control theory, and artificial intelligence to close the loop between perception and purposeful action in energy systems.
Importantly, CEM does not replace the EI; it complements it. The EI provides the infrastructure: the distributed architecture, the communication layers, the packetized energy flows, and the prosumer ecosystem. CEM provides the cognitive governance layer that makes that infrastructure truly intelligent, positioning itself as the missing reasoning engine at the heart of the EI's core server function. Together, EI and CEM form a complete vision: a distributed, software-defined energy network governed not by preset rules or static optimization, but by continuous, goal-directed, adaptive intelligence.

To demonstrate CEM's practical value, this paper grounds it in one of the most operationally demanding EI node environments in modern infrastructure: the smart port. Ports handle over 80\% of the world's freight and are responsible for approximately 3\% of global carbon emissions, with pressure from the International Maritime Organization (IMO) and EU directives intensifying the need for intelligent energy governance \cite{sun2025comprehensive, EU}. Within the port, energy demands are heterogeneous and temporally dynamic:  vessel cold ironing, crane operations, automated guided vehicles (AGVs), and shore-side renewables interact in ways that neither rule-based nor optimization-based systems can govern holistically \cite{teng2022distributed}. This makes the smart port an ideal testbed for CEM: a bounded, high-stakes, multi-resource EI node where cognitive reasoning delivers measurable advantage over conventional approaches. The main contributions of this paper are summarised as follows.

\begin{itemize}
    \item  We formally introduce 
    CEM as a novel conceptual framework, 
    providing the first academic definition of CEM as a goal-directed, 
    self-adaptive energy governance paradigm grounded in cognitive 
    computing principles.

    \item  We position CEM as the cognitive 
    governance layer of the EI, addressing the open  challenge of intelligent energy management in EI-based 
    cyber-physical systems identified in prior work~\cite{hussain2020energy}.

    \item  We distinguish CEM from rule-based and optimization-based EMS paradigms through a structured 
    six-dimensional comparison articulating the fundamental shift from  prescribed to inferred, and from reactive to anticipatory, energy 
    governance.

    \item We develop a simulation-based 
    toy problem grounded in smart port energy management, demonstrating 
    the CEM cognitive loop across a predictive vessel turnaround scenario 
    and quantifying its performance advantages over conventional 
    approaches.
\end{itemize}

The remainder of this paper is structured as follows. Section 2 formally defines CEM and articulates its four-layer cognitive loop. Section 3 distinguishes CEM from conventional energy management paradigms through structured comparison. Section 4 develops the toy problem; a predictive vessel turnaround scenario demonstrating the CEM cognitive loop in action, and Section 5 concludes.
\section{Cognitive Energy Management: Core Concept and Definition}
\subsection{The Intelligence Gap in Energy Systems}
Modern energy management systems (EMS) have evolved considerably over the past two decades. From early rule-based schedulers to sophisticated model predictive controllers and optimization solvers, each generation has brought measurable gains in efficiency and reliability \cite{graa2020review}. Yet across all these generations, a fundamental characteristic has remained unchanged: the system does what it is told. Rules are prescribed by engineers. Objectives are defined at design time. Optimization models assume the world behaves as modeled. When reality deviates, when an unexpected load spike coincides with a forecast error, or when operational context shifts in ways the designer did not anticipate, these systems have no mechanism to understand why or adapt how they respond. As noted in recent assessments of adaptive and intelligent microgrid control, conventional architectures lack the adaptability, predictive capability, and resilience that modern distributed energy environments demand \cite{saeed2026adaptive}.
This gap is not merely a matter of adding more sensors or faster optimization algorithms. It is a structural limitation rooted in the absence of cognition, the capacity to perceive context, reason under uncertainty, learn from experience, and act with purpose. Addressing this gap is the central motivation for introducing CEM.

\subsection{What is Cognitive Energy Management?}
While cognitive computing principles have been applied to enterprise energy management in industrial and patent literature \cite{goparaju2019energy}, and artificial intelligence techniques have been extensively studied for microgrid EMS\cite{ meng2025enhanced, wicaksono2025artificial, hussain2025hybrid, hussain2022benchmarking }, no peer-reviewed work has formally defined CEM as a unified, goal-directed governance framework grounded in the EI architecture. This paper addresses that gap by introducing CEM with the following formal definition:
\textit{A goal-directed energy governance framework in which an energy system continuously perceives its operational context, reasons about current and future states under uncertainty, learns from accumulated experience, and acts with adaptive, anticipatory intelligence to achieve defined energy objectives, without requiring explicit reprogramming for every new condition.}

Three elements of this definition deserve emphasis. First, CEM is goal-directed: it is not merely reactive to stimuli but oriented toward objectives: cost minimization, emissions reduction, and resilience maintenance, which it pursues actively across changing conditions. Second, CEM operates under uncertainty: unlike optimization-based systems that assume known parameters, CEM reasons probabilistically and adapts its plans as new information arrives. Third, CEM is self-improving: through continuous learning, it becomes more effective over time, internalizing patterns that no designer could have fully anticipated.
Importantly, CEM is broader than machine learning applied to energy systems. Machine learning provides pattern recognition from data; cognition provides contextual understanding, goal-directed reasoning, and explainable decision-making \cite{safari2024energy}. A deep reinforcement learning agent can learn to schedule loads efficiently — but it cannot explain why it made a specific decision under a novel operational condition, nor can it reason about a scenario it has never encountered. CEM encompasses ML as one of its enabling mechanisms, but its scope is defined by the cognitive loop it implements, not the algorithms it employs.

\subsection{The Four-Layer Cognitive Loop}
The operational architecture of CEM is organized around a four-layer cognitive loop, inspired by the closed-loop cognitive model of human decision-making \cite{srivani2023cognitive}, and adapted to the energy domain: \\
\textbf{ Layer 1} \underline{Perceive:} The system acquires real-time and contextual data from its operational environment. This includes physical measurements (power flows, state-of-charge, generation output), contextual signals (weather forecasts, market prices, vessel arrival schedules), and historical patterns. Perception in CEM is active, not passive, the system maintains a dynamic situational model of its energy environment, analogous to situational awareness frameworks in power system operations \cite{basu2016situational}.\\
\textbf{Layer 2} \underline{Reason:} Using the situational model, the system reasons about current conditions and anticipated futures. This is where CEM diverges most sharply from conventional EMS: rather than solving a predefined optimization problem, the system evaluates multiple hypotheses, weighs competing objectives, and constructs anticipatory plans under uncertainty. Reasoning in CEM draws on techniques including probabilistic inference, model-based planning, and knowledge representation.\\
\textbf{Layer 3} \underline{Learn:} CEM continuously updates its internal models based on the outcomes of its decisions. When actual outcomes deviate from predictions, a load behaves differently than expected, a renewable source underperforms, and the system revises its understanding. This layer encompasses supervised learning for forecasting, reinforcement learning for policy improvement, and unsupervised learning for pattern discovery \cite{safari2024energy}, \cite{wei2023deep}.\\
\textbf{Layer 4} \underline{Act:} The system executes energy decisions — dispatching storage, scheduling loads, adjusting procurement, routing energy packets, in a manner consistent with its current reasoning and aligned with its goals \cite{hussain2022benchmarking}. Crucially, actions in CEM are not merely feasible; they are intelligent: selected because they best advance the system's goals given its current understanding of the environment.
This four-layer loop operates continuously and cyclically, with each action generating new observations that feed the next perception cycle. The result is an energy management system that does not merely respond to the world as it is, but anticipates the world as it will be.
\subsection{How CEM Connects to the Energy Internet}
The relationship between CEM and the Energy Internet (EI) is structural, not incidental. The EI is defined as a cyber-physical system integrating three tightly coupled layers: energy, communication, and information, coordinated through software-defined control and packetized energy management \cite{hussain2020energy}.
CEM directly maps onto this layered architecture. The \textit{Perceive} layer aligns with the EI information layer by acquiring data from smart meters, sub-energy routers (SERs), and IoT devices. The \textit{Reason} and \textit{Learn} layers operate within the cyber (communication) layer, typically in the core server (CSR), enabling adaptive, goal-oriented decision-making beyond static optimization. The \textit{Act} layer interfaces with the EI energy layer, translating decisions into packetized dispatch actions executed through the SER network \cite{hussain2023heuristic}.
Within this framework, CEM represents a cognitive extension of the IEM concept \cite{hussain2020energy}. While IEM focuses on predefined optimization, CEM introduces reasoning, anticipation, and learning capabilities. The EI both enables this transition—through its rich data infrastructure—and necessitates it due to its inherent system complexity.
This relationship is illustrated conceptually in Fig. 1, which positions CEM as the cognitive intelligence layer at the heart of an EI-based energy node, governing the interaction between physical resources, communication infrastructure, and energy objectives.

\section{CEM Versus Conventional Energy Management Systems}
Energy management has evolved through broadly three paradigm generations, each addressing the shortcomings of its predecessor yet introducing its own structural ceiling. Understanding where each generation breaks down is essential to appreciating what CEM introduces and why it represents a genuine departure rather than an incremental improvement.
First generation, rule-based EMS systems operate through fixed, human-defined thresholds and conditional logic: if state-of-charge falls below X, activate the diesel generator; if grid price exceeds Y, shift load. These systems are transparent and computationally inexpensive, but their intelligence is entirely borrowed from the designer \cite{jamal2024rule}. They cannot handle conditions not anticipated at design time, and their performance degrades precisely when the energy environment becomes most complex;  during high renewable penetration, demand spikes, or novel operational scenarios.
Second generation — Optimization-based EMS systems, including model predictive control (MPC) and mixed-integer linear programming (MILP) approaches, represent a significant advance. By formulating energy management as a mathematical optimization problem, they can handle multiple objectives and operational constraints simultaneously \cite{meng2025enhanced}. However, their fundamental weakness is model dependency: they optimize over a model of the world, and when that model is wrong, as it increasingly is under renewable variability and uncertain load behavior, their solutions are suboptimal or infeasible \cite{lim2025model}. Furthermore, as system complexity grows, their computational burden scales poorly, and they offer no capacity to learn from operational experience.
Third generation, CEM breaks from both preceding paradigms not by doing the same thing better, but by doing something structurally different. 
Table~\ref{tab:cem_comparison} summarizes the key differentiators across six dimensions.

\begin{table*}[t]
\centering
\caption{Comparative Analysis: Rule-Based vs. Optimization-Based vs. Cognitive Energy Management, \cite{jamal2024rule, hussain2022benchmarking, lim2025model, wei2023deep }}
\label{tab:cem_comparison}
\renewcommand{\arraystretch}{1.4}
\begin{tabular}{|p{2.8cm}|p{3.8cm}|p{4.0cm}|p{4.4cm}|}
\hline
\textbf{Dimension} & \textbf{Rule-Based EMS} & \textbf{Optimization-Based EMS} & \textbf{Cognitive EMS (CEM)} \\
\hline
\textbf{Decision Mechanism} 
& Fixed if-then rules 
& Mathematical optimization 
& Goal-directed reasoning under uncertainty \\
\hline
\textbf{Adaptability} 
& None, static rules 
& Limited, model updates required 
& Continuous, learns from every decision cycle \\
\hline
\textbf{Context Awareness} 
& None 
& Partial,  within model assumptions 
& Full, perceives and interprets operational context \\
\hline
\textbf{Uncertainty Handling} 
& None 
& Partial,  robust/stochastic variants 
& Native, probabilistic reasoning built into the loop \\
\hline
\textbf{Learning Capability} 
& None 
& None 
& Core function — improves with experience \\
\hline
\textbf{EI Integration} 
& Peripheral, no CPS awareness 
& Partial,  optimizes over EI resources 
& Native, cognitive loop maps to EI's three layers \\
\hline
\end{tabular}
\vspace{-0.5cm}
\end{table*}
\begin{figure}
    \centering
\includegraphics[width=1\linewidth]{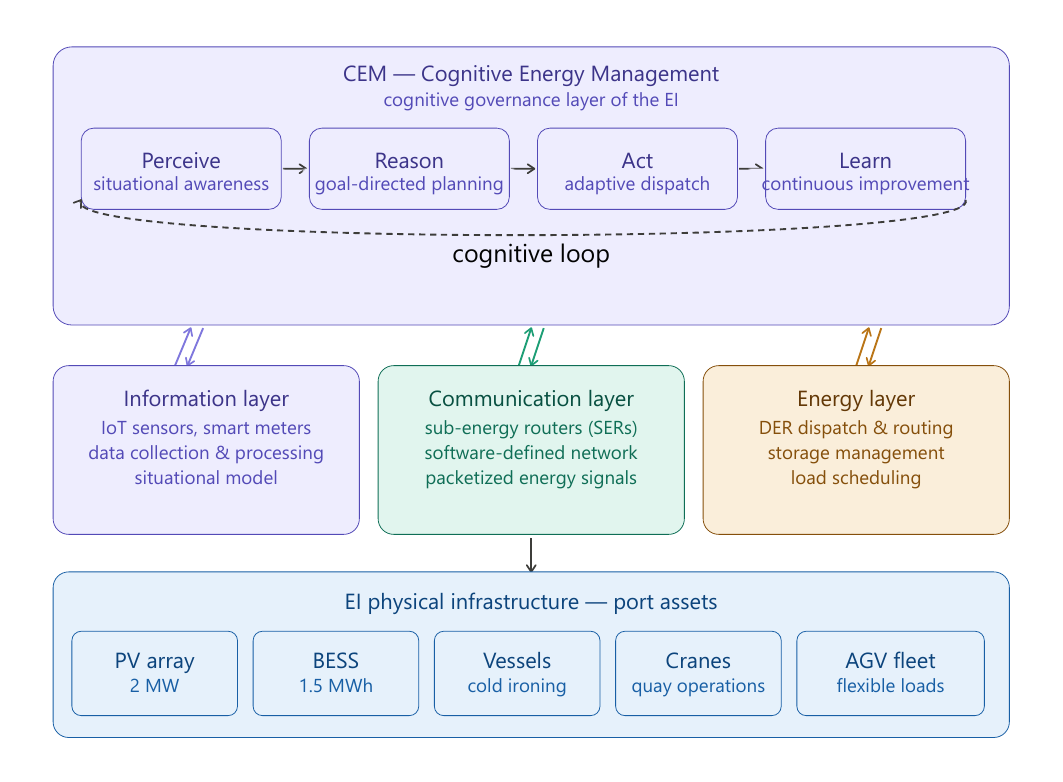}
    \caption{Conceptual overview of CEM in EI architecture}
    \label{fig:CEMvsEI}
    \vspace{-0.3cm}
\end{figure}

\section{ Smart Port Energy Management: A CEM Demonstration}
\subsection{A. The Smart Port as an Energy Internet Node}
Large seaports handle over 80\% of global freight, and under the EU's Alternative Fuels Infrastructure Regulation (AFIR), 100\ shore power provision at core EU ports is mandated by 2030 \cite{lv2025sustainable, fan2023cooperative, hussain2026digital, yang2025assessing}. This electrification wave transforms ports into precisely the multi-resource, multi-network environment that defines an EI node [4]: vessels become mobile prosumers, shore power systems act as energy routers, and battery storage, renewables, and AGV fleets form a heterogeneous DER network that demands intelligent, anticipatory governance \cite{bakar2023electrification}.
\subsection{Scenario Definition}
\begin{figure}
    \centering
    \includegraphics[width=1\linewidth]{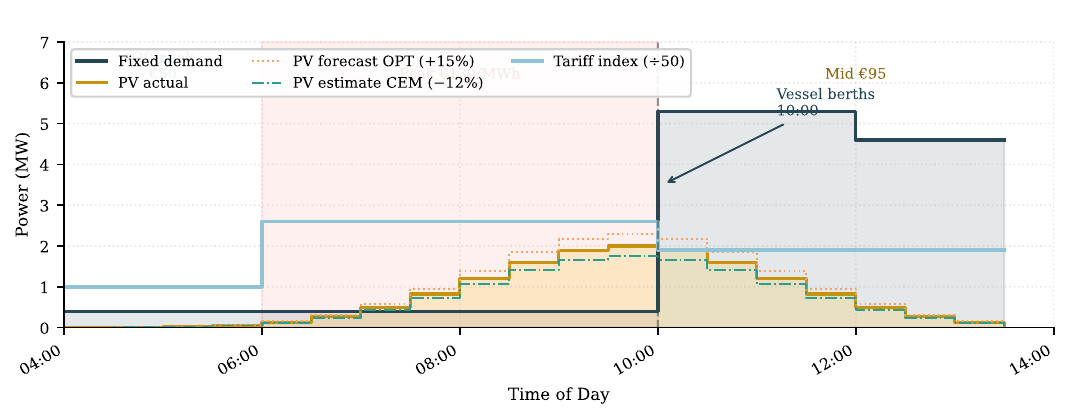}
    \caption{Smart port energy profile over the 10-hour simulation window (04:00–14:00).}
    \label{fig:enery}
    \vspace{-0.5cm}
\end{figure}

We model a 10-hour operational window: 6 hours of pre-berthing horizon (04:00–10:00) followed by a 4-hour vessel turnaround (10:00–14:00). The port operates a 2 MW PV array and a 1.5 MWh BESS (initial SOC = 55\%). Upon berthing, cold ironing imposes a sustained 3.5 MW load; quay crane operations add 1.4 MW (10:00–12:00) tapering to 0.7 MW (12:00–14:00). A 12-vehicle AGV fleet requires 1.2 MWh total charging. Grid tariff follows a time-of-use structure: €50/MWh off-peak (04:00–06:00), €130/MWh peak (06:00–10:00), and €95/MWh mid-peak post-berthing. Fig. \ref{fig:enery} illustrates the full port energy profile, PV generation, and tariff structure.
Three EMS paradigms are simulated against this scenario.
\subsection{CEM Cognitive Loop in Action}
Perceive (T = -6h, 04:00): Upon receiving the vessel arrival confirmation, the CEM system constructs a situational model. It reads (see \ref{fig:emission_combined}a \ref{fig:emission_combined}b) the current BESS SOC (55\%), retrieves solar irradiance forecasts, applying a conservative -12\% adjustment to guard against overestimation, queries the AGV fleet management system (1.2 MWh demand outstanding), and identifies the critical operational tension: peak tariff pricing (€130/MWh) will coincide precisely with maximum PV generation, creating a deceptive surplus that naively-planned systems may over-rely upon.
\subsection{Reason (04:00-06:00)}The CEM system constructs an anticipatory energy plan. It identifies the off-peak window (04:00-06:00) as the optimal period to: (1) complete all AGV charging at €50/MWh before the peak window opens, and (2) pre-charge the BESS aggressively toward SOC$_{max}$, maximising stored energy available for cold ironing. This reasoning is grounded in the EI's packetized energy model; the system treats the BESS as a dispatchable packet reservoir to be filled before the high-demand event \cite{hussain2020energy}.
\subsection{Act (04:00–14:00)} CEM executes its plan. AGV charging is completed entirely within the off-peak window (Fig. \ref{fig:emission_combined123}a). BESS is pre-charged to 95\% SOC by berthing time (Fig. \ref{fig:emission_combined123}b), providing 1.275 MWh of dispatchable energy at €0 marginal cost to offset cold ironing demand. As Fig. \ref{fig:emission_combined}a shows, CEM's grid import profile during the turnaround window is substantially lower than the Rule-Based strategy, as the BESS discharge covers a large fraction of the 3.5 MW cold ironing load.
\subsection{Learn (post-turnaround)} The system compares executed outcomes against its plan. The conservative PV estimate (-12\%) correctly anticipated actual generation. The AGV completion time ran 8 minutes ahead of schedule; the fleet demand model is updated accordingly for future turnarounds.
\subsection{Discussions}
\begin{figure*}[!t]
\centering
\setlength{\abovecaptionskip}{2pt}
\setlength{\belowcaptionskip}{0pt}

\begin{minipage}[b]{0.45\textwidth}
    \centering
    \includegraphics[width=\textwidth, height=0.20\textheight, keepaspectratio]
    {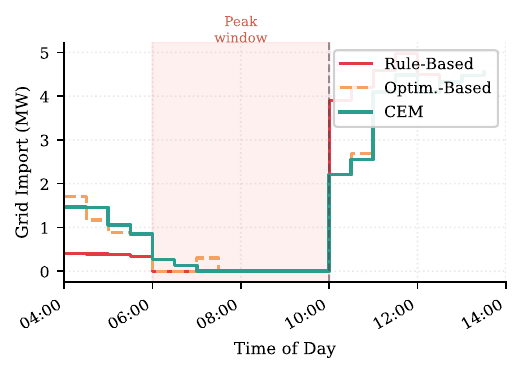}
\end{minipage}
\hspace{0.02\textwidth}
\begin{minipage}[b]{0.45\textwidth}
    \centering
    \includegraphics[width=\textwidth, height=0.20\textheight, keepaspectratio]
    {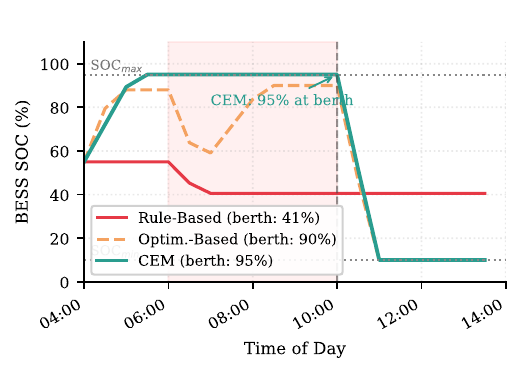}
\end{minipage}
\caption{(a) Grid import profiles for the three EMS strategies over the simulation horizon (b) Battery energy storage system (BESS) state of charge (SOC) trajectories for each strategy.}
\label{fig:emission_combined}
\vspace{-0.4cm}
\end{figure*}

\begin{figure*}[!t]
\centering
\setlength{\abovecaptionskip}{2pt}
\setlength{\belowcaptionskip}{0pt}

\begin{minipage}[b]{0.45\textwidth}
    \centering
    \includegraphics[width=\textwidth, height=0.20\textheight, keepaspectratio]
    {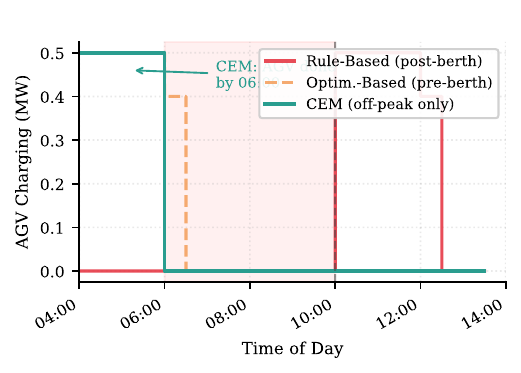} \label{AVG}
\end{minipage}
\hspace{0.02\textwidth}
\begin{minipage}[b]{0.45\textwidth}
    \centering
    \includegraphics[width=\textwidth, height=0.20\textheight, keepaspectratio]
    {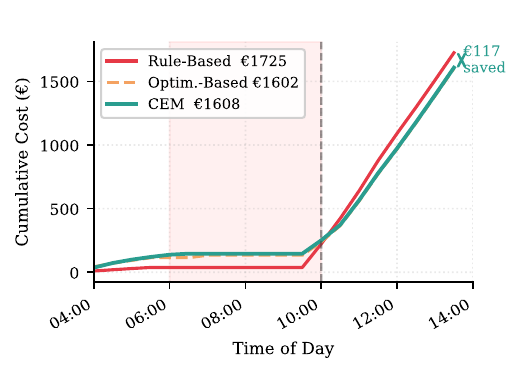} \label{CC}
\end{minipage}
\caption{(a) AGV fleet charging schedules across the three EMS strategies (b) Cumulative energy procurement cost over the 10-hour simulation window.}
\label{fig:emission_combined123}
\vspace{-0.4cm}
\end{figure*}

\begin{figure}
    \centering
    \includegraphics[width=\columnwidth,height=0.50\textheight,keepaspectratio]{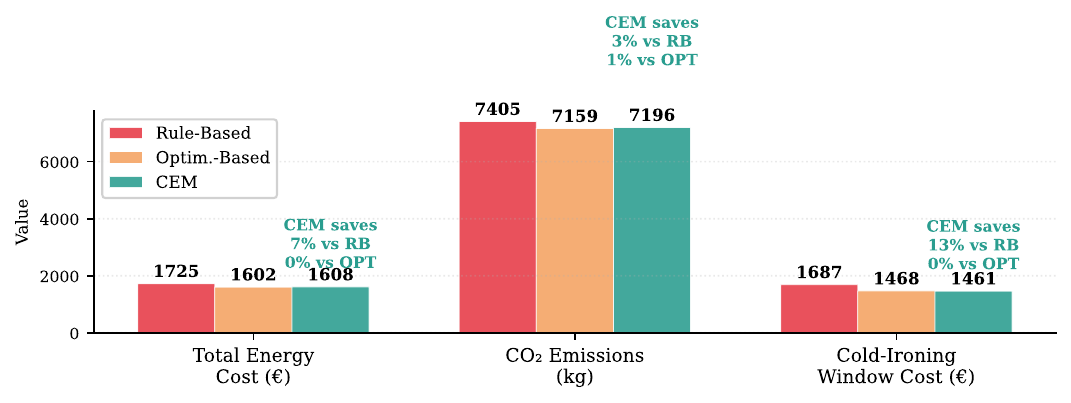}
    \caption{Performance comparison across three key metrics: total energy cost (€), CO$2$ emissions (kg, assuming a grid emission factor of 0.40 kg/kWh), and cold ironing window cost.}
    \label{fig:CEP}
   \vspace{-0.5cm}
\end{figure}

Fig. \ref{fig:enery}–\ref{fig:CEP} present the full simulation results. The rule-Based system, lacking any pre-berthing plan, arrives at berthing with a BESS SOC of only 41\%, insufficient to meaningfully offset cold ironing demand, and schedules AGV charging post-berthing, compounding the load spike. The Optimization-Based system plans ahead using the PV forecast and achieves 90\% SOC at berthing, but its 15\% PV overestimation leaves a residual planning gap. CEM achieves 95\% SOC at berthing through conservative pre-charging, completing AGV scheduling entirely off-peak.
Quantitatively (Fig. \ref{fig:CEP}): CEM reduces total energy cost by 6.8\% versus the Rule-Based strategy (€1608 vs €1725) and cold ironing window cost by 13.4\% (€1461 vs €1687). CO$2$ emissions are reduced by 2.8\% versus Rule-Based. Against the Optimization-Based strategy, CEM delivers comparable total cost while providing greater resilience against PV forecast uncertainty — demonstrated by its higher SOC buffer at berthing.
Table II contextualises these results across the three paradigms.

\section{Conclusion}
This paper introduced Cognitive Energy Management (CEM), a new conceptual framework that addresses a fundamental gap in the Energy Internet paradigm: while EI defines the architecture of next-generation distributed energy systems, it does not resolve how such systems should reason, adapt, and govern energy decisions intelligently. CEM fills this gap by embedding a four-layer cognitive loop: Perceive, Reason, Learn, Act into the energy management process, positioning itself as the cognitive governance layer of EI-based infrastructures. Through structured comparison, we demonstrated that CEM represents a genuine departure from both rule-based and optimization-based paradigms, moving from prescribed to inferred decisions and from reactive to anticipatory energy governance.
To demonstrate CEM's practical value, we developed a simulation-based toy problem grounded in smart port energy management, a compelling EI node environment characterised by heterogeneous loads, prosumer dynamics, and operational intensity. The predictive vessel turnaround scenario showed that CEM reduces total energy cost by 6.8\% and cold ironing window cost by 13.4\% relative to a rule-based strategy, while achieving a BESS state of charge of 95\% at berthing through anticipatory pre-charging,  compared to only 41\% under reactive management.
Several challenges remain open for future work. Integrating CEM with real-time EI communication infrastructure, particularly sub-energy routers and packetized energy dispatch — requires standardised interfaces that do not yet exist at scale \cite{hussain2020energy}. The explainability of CEM's reasoning layer is critical for operator trust in high-stakes port environments, and remains an active research challenge in cognitive computing \cite{safari2024energy}. Furthermore, extending the CEM framework to multi-port or grid-level EI nodes, where prosumer interactions and energy trading dynamics add combinatorial complexity, represents a natural and important next step. 

\begin{table}[t]
\centering
\caption{Simulation Results Summary: Predictive Vessel Turnaround Scenario}
\label{tab:sim_results}
\renewcommand{\arraystretch}{1.1}
\begin{tabular}{|p{1.8cm}|c|c|p{1.4cm}|}
\hline
\textbf{Metric} 
& \textbf{Rule-Based} 
& \textbf{Optim.-Based} 
& \textbf{CEM} \\
\hline
BESS SOC at berthing 
& 41\% 
& 90\% 
& {95\%} \\
\hline
AGV charging window 
& Post-berthing 
& Pre-berthing 
& {Off-peak only} \\
\hline
Total energy cost 
& \euro{}1725 
& \euro{}1602 
& {\euro{}1608} \\
\hline
Cold ironing cost 
& \euro{}1687 
& \euro{}1468 
& {\euro{}1461} \\
\hline
CO\textsubscript{2} emissions 
& 7405 kg 
& 7159 kg 
& {7196 kg} \\
\hline
PV forecast reliance 
& Reactive 
& Optimistic (+15\%) 
& {Conservative (--12\%)} \\
\hline
\end{tabular}
\vspace{-0.5cm}
\end{table}

\bibliographystyle{ieeetr}
\bibliography{Ref}
\end{document}